\documentclass[journal,comsoc]{IEEEtran}
\usepackage[T1]{fontenc}

\usepackage{textcomp}

\usepackage{amsmath,amsfonts,amssymb,amsbsy,amstext}

\usepackage{amsmath}
\usepackage[cmintegrals]{newtxmath}

\usepackage{graphicx}
\usepackage{makecell}
\usepackage{cite}
\usepackage{multirow}
\usepackage{tabularx}
\usepackage{booktabs}

\usepackage{acro}

\NewDocumentCommand{\acro}{m o m o}
{%
	\IfValueTF{#2}{%
		\IfValueTF{#4}{%
			\DeclareAcronym{#1}{short={#2},long={#3},#4}
		}{%
		  \DeclareAcronym{#1}{short={#2},long={#3}}
	  }
  }{%
    \IfValueTF{#4}{%
	    \DeclareAcronym{#1}{short={#1},long={#3},#4}
    }{%
      \DeclareAcronym{#1}{short={#1},long={#3}}
    }
  }
}

\acro{1G}{First Generation}
\acro{2G}{Second Generation}
\acro{3G}{Third Generation}
\acro{4G}{Fourth Generation}
\acro{5G}{fifth generation}
\acro{B5G}{beyond \ac{5G}}
\acro{3GPP}{3rd Generation Partnership Project}
\acro{3GPP2}{Third Generation Partnership Project 2}
\acro{ADMM}{alternating direction method of multipliers}
\acro{BB}{branch and bound}
\acro{BER}{bit error rate}
\acro{BS}{base station}
\acro{CDF}{cumulative distribution function}
\acro{CFN}{channel Frobenius norm}
\acro{CSI}{channel state information}
\acro{CoMP}{coordinated multipoint}
\acro{DCP}{difference of convex functions program}
\acro{eMBB}{enhanced mobile broadband}
\acro{GTEL}{Wireless Telecommunications Research Group}
\acro{GUB}{global upper bound}
\acro{GLB}{global lower bound}
\acro{HARQ}{hybrid automatic repeat request}
\acro{HMSINR}{highest minimum \ac{SINR}}
\acro{HWCR}{highest \ac{WCR}}
\acro{HWSR}{highest weighted sum rate}
\acro{IBC}{interference broadcast channel}
\acro{IC}{interference channel}
\acro{IoT}{Internet of Things}
\acro{KKT}{Karush-Kuhn-Tucker}
\acro{LB}{lower bound}
\acro{LOS}{line of sight}
\acro{LP}{linear programming}
\acro{mMTC}{massive machine-type communications}
\acro{MILP}{mixed integer linear programming}
\acro{MIMO}{multiple input multiple output}
\acro{MIMOIBC}{\ac{MIMO} \ac{IBC}}
\acro{MINLP}{mixed integer nonlinear programming}
\acro{MIP}{mixed integer programming}
\acro{MISO}{multiple input single output}
\acro{MMSE}{minimum mean squared error}
\acro{MRC}{maximum ratio combining}
\acro{MRT}{maximum ratio transmission}
\acro{MSE}{mean squared error}
\acro{MU-MIMO}{multi-user multiple input multiple output}
\acro{MU}{multi-user}
\acro{NLP}{nonlinear programming}
\acro{NLOS}{non-line of sight}
\acro{NP}{non-polynomial time}
\acro{NR}{New Radio}
\acro{RRV}{rate-relaxation variable}
\acro{RRH}{remote radio head}
\acro{RSU}{road side unit}
\acro{QoS}{quality of service}
\acro{SCA}{successive convex approximation}
\acro{SINR}{signal to interference-plus-noise ratio}
\acro{SISO}{single input single output}
\acro{SIMO}{single input multiple output}
\acro{SNR}{Signal to Noise Ratio}
\acro{SOCP}{second-order cone programming}
\acro{TTI}{transmission time interval}
\acro{TU}{typical urban}
\acro{UE}{user equipment}
\acro{UB}{upper bound}
\acro{UAV}{unmanned aerial vehicles}
\acro{URLLC}{ultra-reliable low-latency communications}
\acro{V2X}{vehicle-to-everything}
\acro{WINNER}{Wireless World Initiative New Radio}
\acro{WMMSE}{weighted minimum mean squared error}
\acro{WCR}{weighted common rate}
\acro{WP}{work package}
\acro{ZF}{zero-forcing}
\acro{ZMCSCG}{Zero Mean Circularly Symmetric Complex Gaussian}

\usepackage{siunitx}
\usepackage{datetime}
\usepackage{url}

\usepackage{subfig}

\usepackage[hidelinks=true]{hyperref}

\newcommand{\FigRef}[2][]{Fig.#1~\ref{#2}}

\usepackage{color}

\begin{document}

\title{Decentralized Joint Transceiver Design and QoS Management in 5G and Beyond Systems}
\title{Decentralized Joint {Beamforming, User Scheduling} and QoS Management in 5G and Beyond Systems}
\author{ Roberto P. Antonioli, G\'abor Fodor, Pablo Soldati and Tarcisio F. Maciel
	\thanks{This work was supported by Ericsson Research, Technical Cooperation contract UFC.46 (EDB/NAIVE). This study was financed in part by the Coordena\c{c}\~{a}o de Aperfei\c{c}oamento de Pessoal de N\'{\i}vel Superior - Brasil (CAPES) - Finance Code 001, CNPq and FUNCAP. The authors acknowledge the assistance of Iran M. Braga Jr. for drawing Figure 1.}
	\thanks{Roberto P. Antonioli and Tarcisio F. Maciel are with the  Wireless Telecom Research Group (GTEL), Federal University of Cear\'a, Fortaleza, Brazil. E-mails: \{antonioli, maciel\}@gtel.ufc.br. G\'abor Fodor and Pablo Soldati are with  Ericsson Research, Stockholm, Sweden. E-mails: \{gabor.fodor, pablo.soldati\}@ericsson.com. G\'abor Fodor is also with the KTH Royal Institute of Technology, Stockholm, Sweden. E-mail: gaborf@kth.se.}
	}

\maketitle

\begin{abstract}
Fifth generation cellular systems support a broad range of services, 
including mobile broadband, critical and massive Internet of Things
and are used in a variety of scenarios.
In many of these scenarios, the main challenge is maintaining high throughput 
and ensuring proper quality of service (QoS) in irregular topologies.
{In multiple input multiple output systems,} 
this challenge translates to designing linear {transmit and receive beamformers} that 
maximize the system throughput and manage QoS constraints.
In this paper, we argue that this basic design task {in 5G and beyond systems} must be extended such
that {beamforming} design and user scheduling are managed jointly.
Specifically, we propose a fully decentralized joint {beamforming} design and
user scheduling algorithm that manages QoS.
A novel feature of this scheme is its ability to reduce the initial rate requirements
in case of infeasibility.
{By means of simulations that model contemporary 5G scenarios}, 
we show that the proposed decentralized scheme outperforms benchmarking 
algorithms that do not support minimum rate requirements and 
previously proposed algorithms that support QoS requirements.
\end{abstract}


\IEEEpeerreviewmaketitle

\acresetall

\section{Introduction}
\label{Sec:Intro}
Current and emerging wireless communications technologies
support a wide variety of applications and services, ranging from mobile broadband {to} ultra-reliable low-latency,
{broadband Internet of Things}, including
{vehicular communications, advanced driver assistance systems, communication between \acp{UAV},
factory automation, high-resolution gaming etc.~\cite{Guey2015, Lohmar2019}.}
Such a broad range of application scenarios and services poses an equally broad set of design requirements
and \ac{QoS} demands {in terms of} latency, reliability, energy and spectral efficiency.
For example, remote driving -- supported by the fifth generation (5G) of wireless networks -- requires less than 10 msec latency and 99.999\%
reliability over the radio interface, while advanced driving assistance systems
employing augmented vehicular reality technology require several 100 Mb/s in the
downlink \cite{Qiu:17,Ashraf2020,3gpp.38.885}.

\begin{figure}[t!]
	\centering
	\includegraphics[trim={1.8cm 1.9cm 1.8cm 1.2cm},clip,scale=0.54]{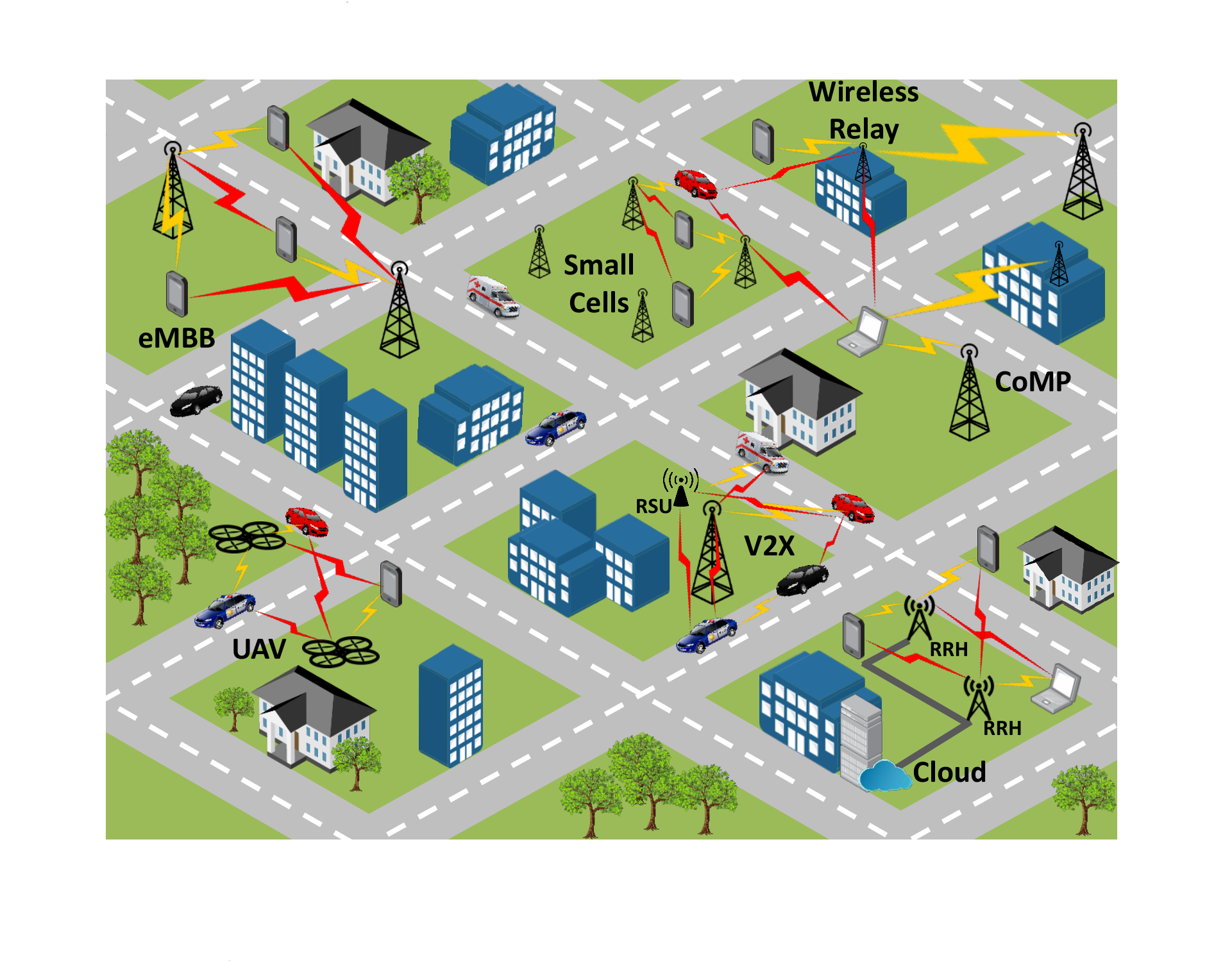}
	\caption{Examples of deployment scenarios involving various wireless nodes and links
             that can be conveniently categorized as MIMO {interference broadcast channel} scenarios.}
	\label{fig:intro:scenarios}
\end{figure}

Addressing all these design challenges necessitates a variety of solutions in terms of enhanced deployment scenarios and more advanced transmission and reception technologies. 
Network deployments are becoming increasingly dense and heterogeneous, with enhanced cellular base stations being co-deployed with wireless relay nodes, remote radio heads,
wireless roadside units, drone-mounted base stations, multiple transmit and reception points and cell-free configurations~\cite{Atzeni2021}. 
These solutions will support an ever-growing number of connections to a large variety of user equipment, ranging from typical mobile handsets to connected vehicles, \acp{UAV}, low-power wireless sensors and actuators, etc., as illustrated in~\FigRef{fig:intro:scenarios}. 
The resulting system is typically interference-limited~\cite{Soret2018}.
On the other hand, starting from the fourth-generation (4G) LTE systems, communication networks are featuring an increasing number of antennas, both at network nodes and user equipments. Such systems exploit advanced \ac{MIMO} transmission and reception (transceiver) algorithms to capitalize on multi-path propagation diversity and spatial multiplexing and increase spectral efficiency.
{Such deployment environments} can be characterized as systems operating using a \ac{MIMO} \ac{IBC}.

{To address the needs of emerging communication technologies operating in \ac{MIMO} \ac{IBC} scenarios,
we argue that two design features are of paramount importance: (a)} joint procedures of transceiver {beamforming} design and link scheduling for QoS management capable of ensuring that the links either meet the respective QoS requirements or suffer only a minor QoS degradation; and (b) decentralized solutions to suit the distributed nature of modern communication systems.
{These two features are of high interest in several \ac{MIMO} \ac{IBC} systems being currently investigated by \ac{3GPP} {and illustrated in~\FigRef{fig:intro:scenarios}}, such as \ac{NR} \ac{V2X}~\cite{3gpp.38.885}, \ac{UAV}~\cite{3gpp.22.125} as well as general \ac{IoT} scenarios~\cite{Petrov2019}.}

{While the literature on \ac{MIMO} \ac{IBC} systems provides an abundance of solutions for the transceiver design, only a handful of works have addressed link scheduling in \ac{MIMO}~\ac{IBC}} and even fewer have considered \ac{QoS} requirements (for example, in terms of \ac{SINR}, date rate, tolerable latency, and energy efficiency).
{This niche of works, however, either focuses entirely on the transceiver design
(assuming feasible QoS requirements) or relies on centralized solutions to handle link scheduling and QoS management}.
{Nevertheless, the \ac{MIMO}~\ac{IBC} arises in many broadband Internet of Things scenarios for which
the existing centralized solutions might not be applicable.
Concerning massive \ac{MIMO} deployments, it could be argued that
when perfect or high-quality \ac{CSI} is available, the links become orthogonal,
rendering the joint transceiver design and link scheduling superfluous.
However, many real-life deployments are subject to \ac{CSI} errors and have limited number of antennas{~\cite{3gpp.38.885,Petrov2019}}.
Therefore, it is often necessary to develop solutions for decentralized transceiver {beamforming} design
and link scheduling for \ac{QoS} {management} in \ac{MIMO}~\ac{IBC} deployments.}

To fill this gap, this work presents a novel concept {that jointly tackles the beamforming design, user scheduling and \ac{QoS} management}
for \ac{QoS}-constrained \ac{MIMO}~\ac{IBC} systems. 
{Jointly addressing these three problems is novel and highly relevant to \ac{5G} and beyond \ac{MIMO} systems.
Moreover, the proposed concept is highly different from previous works, which have addressed the problems of beamforming design, user scheduling and \ac{QoS} management separately or by proposing complex and centralized solutions to address some combination of the above three problems. }
{The key feature of our approach is} an optimization mechanism for QoS management that, {unlike related works in the literature,} manages \ac{QoS} in a decentralized fashion {and handles infeasible instances of \ac{QoS} requirements by allowing (and optimizing)} a \ac{QoS} degradation for some links, {which eventually results in some link deactivation when inevitable}.
We discuss the algorithmic and signaling aspects of implementing the proposed solution in practical systems{, which are aspects often ignored in previous works.}
{Finally, }simulations show the advantages of the proposed joint transceiver {beamforming} design and link scheduling compared to existing solutions.

\section{{Current {Beamforming Design, User Scheduling} and QoS Management in MIMO Systems}}
\label{Sec:RelatedWorks}

The importance of transceiver {beamforming} design and link scheduling for the \ac{MIMO} \ac{IBC} has been recognized for long and a rich literature of solution approaches exists.
We highlight the main lines of research and discuss outstanding issues related to implementing some of the proposed solutions in contemporary \ac{MIMO}~\ac{IBC} scenarios.

{Considering single-antenna systems, the} problem of link scheduling with \ac{QoS} constraints has been widely studied in the literature {and several approaches have been implemented in practice~\cite{Capozzi2013}}.
{In such single-antenna systems, the link scheduling task is mainly conducted by selecting appropriate power levels depending on the targeted objective.}
These solutions are not applicable in today's \ac{MIMO} systems, because when the transmitters and receivers are equipped with multiple antennas, proper transmit and receive vectors that facilitate beamforming and spatial-division multiplexing must be designed.

Transceiver {beamforming} design and optimization, on the other hand, has been extensively studied for \ac{MIMO} \ac{IBC} systems{, some of which might be of interest for application in 5G and beyond emerging radio technologies}.
To this end, a widely pursued performance metric is the maximization of the total system weighted sum-rate.
Such metric is of high interest for network operators as the simple choice of its weights allows to steer the allocation of radio resources towards different goals, e.g., from maximizing the overall system throughput to ensuring some form of fairness among users.
With this objective in mind, the authors in Shi \textit{et~al.}~\cite{Shi2011} proposed an iterative and decentralized transceiver design mechanism based on the \ac{WMMSE} approach.
While Shi \textit{et al.}~\cite{Shi2011} has stimulated more research in this area, this approach is limited to transceiver design for unconstrained sum-rate maximization{, i.e., the authors ignored \ac{QoS} management aspects.}

In {5G and beyond wireless communications} systems, however, many services {will} demand certain \ac{QoS} from the underlying communication links.
From an optimization {prospective, adding} \ac{QoS} constraints {may lead to} feasibility issues in interference-limited systems, {which per se is a non-trivial challenge}.
Transceiver design considering \ac{QoS} demands for the weighted sum-rate maximization was considered in~\cite{Kaleva2016}.
{This work, however, only addresses the transceiver design aspect requiring} that a feasible set of links was previously selected {prior to} executing the transceiver {optimization}{, meaning that the authors ignored user scheduling aspects.}

\begin{figure}[!t]
	\centering
	\subfloat[User set reduction approach, where user-2 is not scheduled while the remaining users simultaneously have their QoS demands met.]{\includegraphics[scale=0.32]{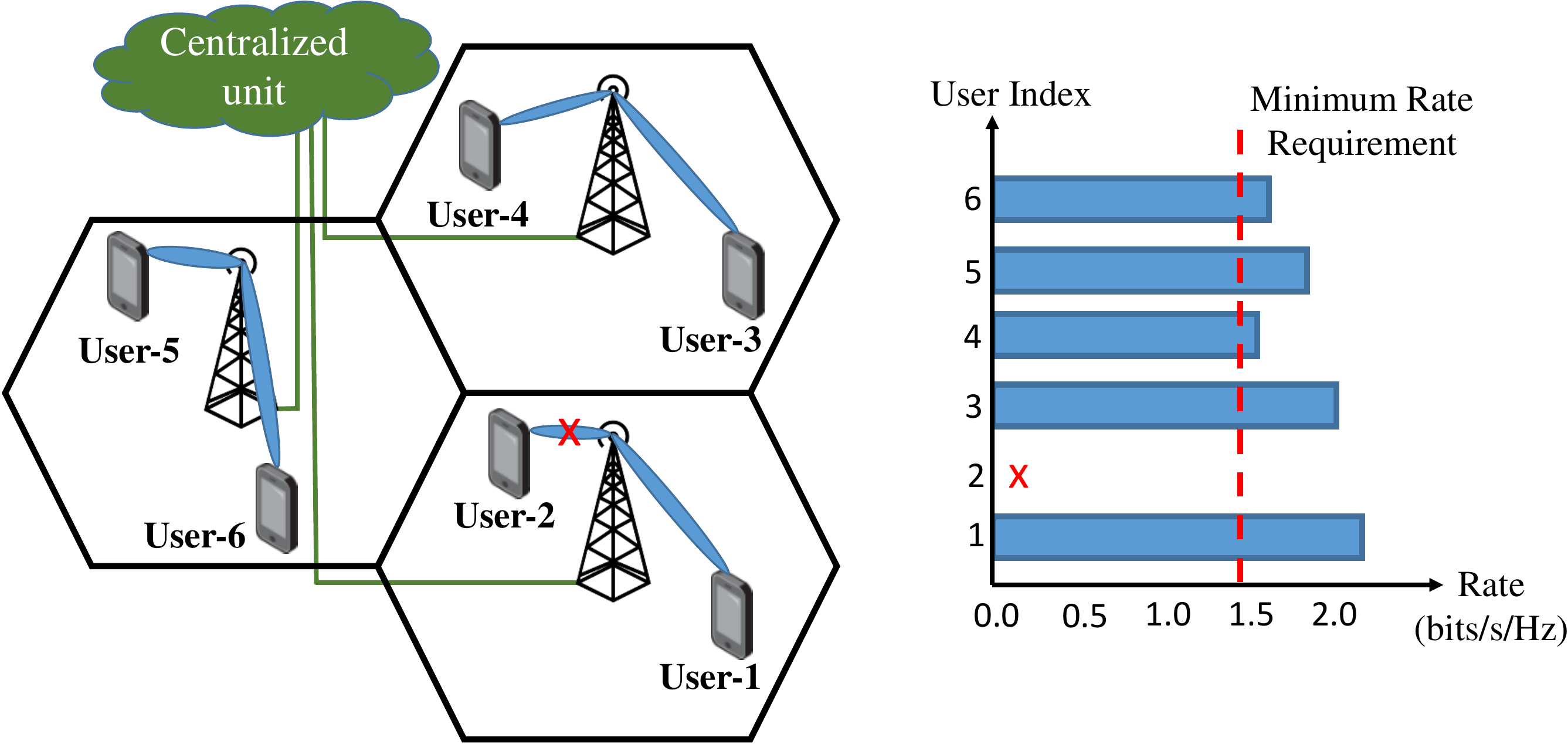}%
		\label{fig:usersetreduction}}
	
	\subfloat[Relaxation of QoS constraints approach, where a relaxed common QoS requirement is set so that every user meets the new requirement.]{\includegraphics[scale=0.32]{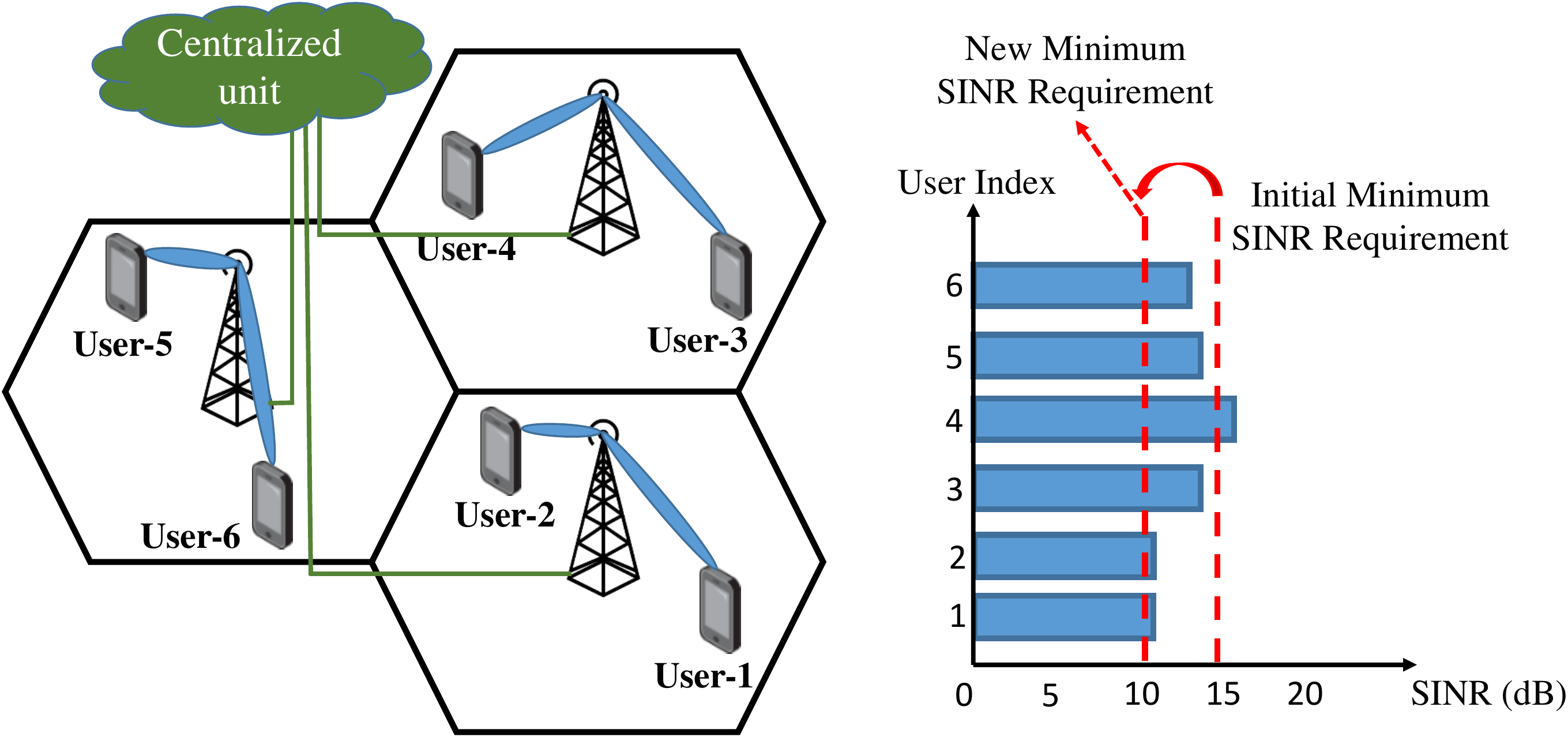}%
		\label{fig:qosrelaxationapproach}}	
	\caption{Illustration of existing approaches for link scheduling and transceiver design with QoS management considering a cellular MIMO IBC scenario.}
	\label{fig:overview:sched}
\end{figure}

{The feasibility issues brought by per-user QoS requirements can naturally be seen as part of a scheduling problem, i.e., finding a subset of users that can be co-scheduled in the same radio resource so as to optimize the desired performance metric while satisfying the users' respective QoS demands.}
In the context of \ac{MIMO} \ac{IBC} systems with per-user \ac{QoS} requirements, {joint} transceiver design and link scheduling with \ac{QoS} management consists of optimizing the transceiver beamforming vectors while selecting a set of users for which the respective \ac{QoS} demands can be simultaneously fulfilled or suffer only a minor \ac{QoS} degradation.
These two scheduling approaches are referred in the literature as the user set reduction approach and the relaxation of \ac{QoS} constraints approach, respectively, which are illustrated in~\FigRef{fig:overview:sched}.

\FigRef{fig:usersetreduction} {depicts an example of} user set reduction {with \ac{QoS} demands in terms of minimum rate requirements. User-2 in this example is} deactivated so that the \ac{QoS} demands of the remaining subset of users {can be} simultaneously fulfilled.
\FigRef{fig:qosrelaxationapproach}, illustrates {an example of} relaxation of \ac{QoS} constraints  {in terms of minimum rate requirement}. {This approach} relaxes the \ac{QoS} requirements {to find a common minimum \ac{QoS} requirement that can be met by} the users simultaneously.

{Solutions available in the literature for transceiver design and link scheduling in MIMO IBC systems with QoS requirements, based on either of these two approaches for QoS management, consider the aid of a centralized unit to perform at least the link scheduling}.
{Centralized user set reduction (cf. \FigRef{fig:usersetreduction}) is proposed, for instance, in~\cite{Yu2012} for \ac{MISO}~\ac{IBC} systems} to deactivate some links so as to satisfy the minimum \ac{SINR} requirements for the remaining active links while minimizing the power expenditure in the system.
{In the context of} MIMO IBC, the authors in~\cite{Antonioli2019} proposed link scheduling and transceiver design
solutions for maximizing the sum-rate under minimum rate constraints,
where a centralized unit is employed to perform the link scheduling,
while the transceiver design is computed in a decentralized fashion.
Regarding the relaxation of \ac{QoS} constraints approach shown in~\FigRef{fig:qosrelaxationapproach},
a centralized unit was used in~\cite{Tolli2008} to compute a minimum \ac{QoS} requirement feasible for all users in terms
of maximum minimum weighted common rate.

The use of a centralized unit, however, might not be feasible nor desirable in many {5G and beyond} \ac{MIMO} \ac{IBC} scenarios, such as \ac{V2X} and \ac{UAV} scenarios,
implying that the existing solutions in~\cite{Yu2012,Antonioli2019,Tolli2008}
cannot be applied to modern scenarios.
Therefore, {designing} decentralized solutions for joint link scheduling
and transceiver {beamforming} design considering \ac{QoS}
management in {5G and beyond emerging radio technologies operating in} the \ac{MIMO} \ac{IBC} remains an open problem.

\section{{Decentralizing {Joint Beamforming, User Scheduling} and QoS Management in 5G and Beyond MIMO Systems}}

{Meeting the requirements imposed by the stringent and heterogeneous demands 
of new communication services 
requires advanced \ac{QoS} management solutions. 
To this end, we propose a novel concept to address \ac{QoS} management in \ac{MIMO} \ac{IBC} 
systems based on jointly computing the transceiver beamforming vectors and link scheduling.}
The proposed solution {has}
two main advantages:
(i) the computations are executed in a decentralized fashion,
making it {suitable in} 
{5G and beyond systems, conveniently modelled as} 
\ac{MIMO} \ac{IBC},
and 
(ii) the solution employs {an optimization framework} that {generalizes and combines} 
existing scheduling approaches {for QoS management}. 
{Specifically, it combines
user set reduction and relaxation of \ac{QoS} requirements} into a single mechanism that handles infeasible QoS instances by either scheduling 
links with reduced \ac{QoS} requirements or deactivating some links. 

{The first aspect to be highlighted 
is that -- as opposed to {alternative methods} -- our solution {is fully distributed, thus} it does not require a centralized unit.
This characteristic is extremely important in {5G and beyond wireless communications} systems
that are inherently distributed and cannot rely on a centralized unit.}
Nevertheless, the proposed solution requires some communication between transmitters
and receivers using backhaul links (when available) or using over-the-air signaling schemes, as for instance~\cite{Tolli2019}, which will be described in the next section.
{It is worth highlighting that - as opposed to existing solutions - the proposed solution jointly handles the beamforming design, user scheduling and \ac{QoS} management problems, which are relevant aspects need by \ac{5G} and beyond \ac{MIMO} systems.} 

\subsection{{System Model and High-Level Description of the Proposed Optimization Problem}}

{We consider a general setup of \ac{5G} and beyond systems, conveniently modeled as a \ac{MIMO} \ac{IBC} system, in which $B$ transmitting nodes equipped with $N_\text{T}$ antennas serve in total $U$ multi-antenna receivers each equipped with $N_\text{R}$ antennas.
All transmitting nodes operate over a common frequency channel and serve the respective users with linear transmit beamforming.
$S_u$ denotes a fixed number of spatial streams allocated to receiver $u$ and we indicate the $s$th stream of receiver $u$ by the pair $(u,s)$.}

{The downlink signal received by receiver $u$ over spatial stream $s$ {is} expressed as}
\begin{equation}
{\textbf{y}_{u,s} = \textbf{H}_{b_u,u} \textbf{m}_{u,s} x_{u,s}
+ \sum_{i=1}^{U} \hspace{-2ex} \mathop{\sum_{j=1,}}_{(i,j)\neq(u,s)}^{S_i} \hspace{-2ex} \textbf{H}_{b_i,u}\textbf{m}_{i,j} x_{i,j} + \textbf{n}_{u}},
\end{equation}
{where $\textbf{H}_{b_i,u} \in \mathbb{C}^{N_\text{R} \times N_\text{T}}$ is the channel matrix between receiver $u$ and transmitter $b$ serving receiver $i$,
$\textbf{m}_{u,s} \in \mathbb{C}^{N_\text{T}}$ is the transmit beamforming vector of the corresponding data stream,
$x_{u,s}$ is the mutually independent transmitted data symbol with $\mathbb{E} \left[ |x_{u,s}|^2 \right] = 1$ and
$\textbf{n}_{u} \in \mathbb{C}^{N_\text{R}} \sim CN(0,\sigma_u^2)$ is the noise at user ${u}$.
User $u$ decodes the signal $\textbf{y}_{u,s}$ via a unit norm receive beamformer~$\textbf{w}_{u,s}\in\mathbb{C}^{N_\text{R}}$.
The \ac{SINR} for stream $s$ of user $u$ is given by}
\begin{equation}
\label{eq:SINR}
{\Gamma_{u,s} = \dfrac{|\textbf{w}_{u,s}^H \textbf{H}_{b_u,u} \textbf{m}_{u,s} |^2}{ \sum\limits_{i=1}^{U} \mathop{\sum\limits_{j=1,}}\limits_{(i,j)\neq(u,s)}^{S_i} | \textbf{w}_{u,s}^H \textbf{H}_{b_i,u}\textbf{m}_{i,j} |^2 + \sigma_u^2\lVert\textbf{w}_{u,s}\rVert^2}.}
\end{equation}

The proposed concept relies on formulating an optimization problem wherein the individual \ac{QoS} requirements are modeled as constraints to which an optimization variable is added to enable the optimization of \ac{QoS} management in case of infeasible instances of the problem.
Such an additional optimization variable is used to control the \ac{QoS} degradation incurred to the users in infeasible scenarios, while a penalty function of this variable is subtracted in the objective function to diminish the suffered \ac{QoS} degradation.
{Mathematically, the proposed optimization problem can be formulated as:}
\begin{align}\hspace{-0.4cm}
{\begin{array} {lll}
\underset{\mathbf{w}_{u,s}, \mathbf{m}_{u,s}, d_u}{\text{maximize}} & \sum\limits_{u=1}^U  \beta_{u} \left( g(\Gamma_{u,s}) - f(d_u)\right)& \\
\mbox{subject to} & \sum\limits_{u\in\mathcal{U}_b} \sum\limits_{s=1}^{S_u} \lVert\mathbf{m}_{u,s}\rVert^2 \leq P_{b}, & \forall b,\\
& \text{QoS requirement constraint}, & \forall u,
\end{array} \label{PROB:MAX_RATE_SLACK}}
\end{align}
{where $\beta_{u}>0$ models the priority weight and \ac{QoS} requirement of receiver $u$ and $P_b$ is the power budget of transmitter $b$.
The \ac{QoS} requirements constraints can be modeled as $g(\Gamma_{u,s}) \geq \text{QoS}_u - d_u$, in case there is a minimum per-receiver \ac{QoS} requirement (e.g., rate or \ac{SINR}) or as $g(\Gamma_{u,s}) \leq \text{QoS}_u + d_u$, in case there is a maximum per-receiver \ac{QoS} requirement (e.g., packet latency or outage probability), where $\text{QoS}_u$ models the \ac{QoS} requirement of receiver $u$.
Thus, besides the transmit-receive beamforming variables $\{\textbf{m}_{u,s}, \textbf{w}_{u,s}\}_{\forall (u,s)}$, the optimization is conducted over the variables~$\{d_u\}_{\forall u}$, which allows the \ac{QoS} constraints to be relaxed.
Therefore, $d_u$ is a \ac{QoS} relaxation variable.
Finally, one should have constraints restricting the domain of $d_u$, where the formulation of such constraints varies depending on the considered \ac{QoS} constraints.
For instance, one could have $d_u(d_u-\text{QoS}_u) \leq 0$ in case there are minimum per-receiver \ac{QoS} requirements.
These additional optimization variables $d_u$ are used to control the \ac{QoS} degradation incurred to the users in infeasible scenarios, while a penalty function of this variable is subtracted in the objective function to diminish the suffered \ac{QoS} degradation.}

{One example of the application of problem~\eqref{PROB:MAX_RATE_SLACK} is to consider a penalized sum-rate maximization problem subject to minimum rate constraints, in which $g(\Gamma_{u,s})$ models the rate of receiver $u$, the \ac{QoS} constraints are expressed in terms of minimum rate constraints, and, thus, the $d_u$ variables are rate relaxation variables. Nevertheless, it} is worth noting that our concept is applicable to a broad range of \ac{QoS} requirements, such as minimum rate or \ac{SINR} requirements as well as maximum packet latency. 
Moreover, our concept supports many types of optimization of a key performance indicator, such as maximizing the total system sum-rate or minimizing the power expenditure in the system.

\subsection{{Illustrative Examples of the Proposed Concept}}

\begin{figure}[!b]
	\centering
	\subfloat[Proposed solution only relaxes the QoS requirement of user-2 while meeting the rate requirements of the remaining users. In this case, it is not necessary to not schedule {(i.e. deactivate)} any user.]{\includegraphics[scale=0.32]{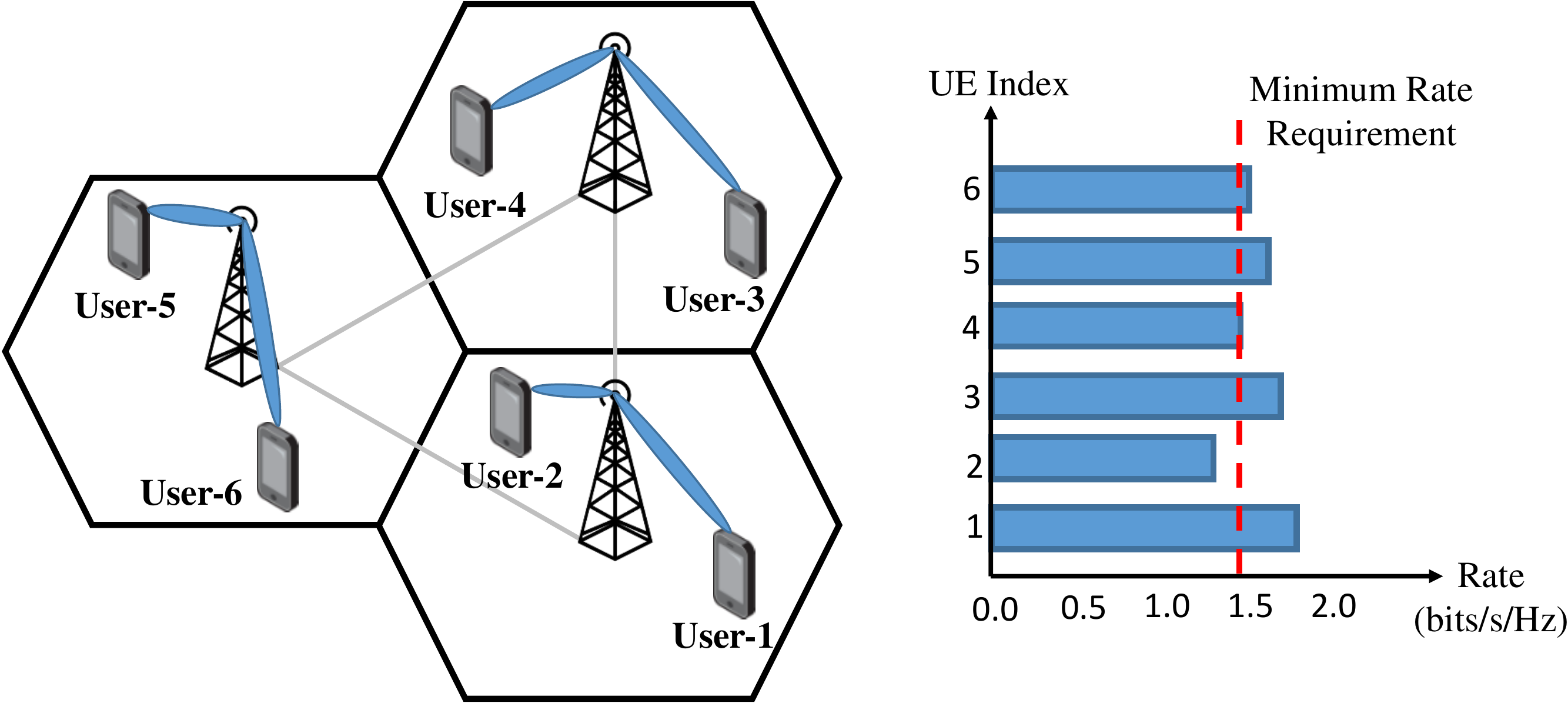}%
		\label{fig:proposedRelax}}
	\hspace{1cm}	
	\subfloat[Proposed solution performs a joint relaxation of QoS constraints and user set reduction, where user-1 is not scheduled and the QoS requirement of user-2 is relaxed.]{\includegraphics[scale=0.32]{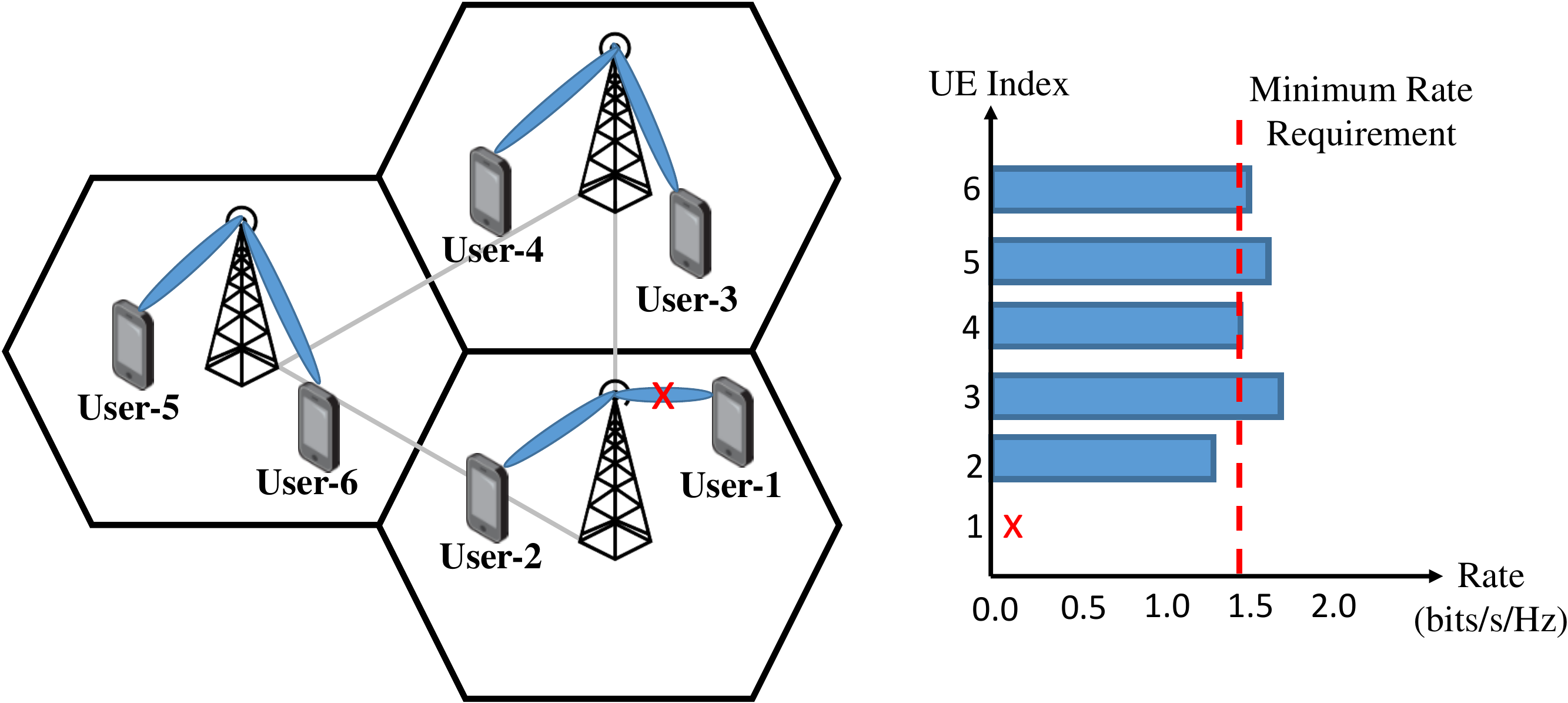}%
		\label{fig:proposedRelaxDeact}}	
	\caption{Rate allocation of proposed solution for joint link scheduling and transceiver design with QoS management considering cellular MIMO IBC scenarios.}
	\label{fig:proposed}
\end{figure}

\FigRef{fig:proposed} exemplifies the main {features} of the proposed {solution concept} considering a cellular \ac{MIMO}~\ac{IBC} scenario with QoS requirements in terms of per-user {minimum rate requirements}.
{Even though we use a traditional cellular \ac{MIMO}~\ac{IBC} scenario 
as example in~\FigRef{fig:proposed}, the proposed solution is applicable to any 5G scenario 
that operates in a \ac{MIMO} \ac{IBC}, such as the ones illustrated in~\FigRef{fig:intro:scenarios}.}
{The first feature, showed in \FigRef{fig:proposedRelax}, is the ability of optimizing the degradation of the QoS requirements of individual users in case of infeasible instances of QoS requirements. As such, our approach generalizes the} relaxation of \ac{QoS} constraints concept  (cf. Section~\ref{Sec:RelatedWorks}). In particular, \FigRef{fig:proposedRelax} {shows an example wherein} it is not possible to design a set of transmit and receiver beamforming vectors while meeting the minimum \ac{QoS} requirement for all users.
{By allowing to optimize the QoS demands of individual users, the proposed mechanism returns a solution where only} the \ac{QoS} requirement of user-2 is reduced, while meeting the \ac{QoS} demands of the remaining users.
{This is different from} existing solutions based on relaxation of \ac{QoS} constraints (e.g.,~\cite{Tolli2008}) {which strive to find a minimum common \ac{QoS} requirement} that can be supported by all users (as shown, e.g. in~\FigRef{fig:qosrelaxationapproach}).
On the other hand, {this solution is also different from} existing mechanisms that adopt the user set reduction approach, {which} would allocate zero rate to one user (possibly user-2) to meet the \ac{QoS} demands of the remaining users, thus not fully optimizing the \ac{QoS} management for that scenario.
Therefore, our solution, in fact,
{allows some controlled \ac{QoS} degradation} 
aiming at {degrading} as {little} as possible
the per-user \ac{QoS} provisioning in infeasible scenarios.

Even though the rate provided to user-2 is below its minimum rate requirement, it can still be useful {in many} services consumed by that user.
For example, consider the situation where all users demand minimum rates for streaming a 4K video, but it is not possible to {support all the demands simultaneously}.
In this situation, the proposed solution computes a rate allocation that allows some users to watch the video in a lower resolution (e.g., 1080p), such as user-2 in~\FigRef{fig:proposedRelax}, while the remaining users watch the video in 4K.
Besides that, since our solution is executed considering a single time-frequency resource,
user-2 can still be served in other time-frequency resources to meet its minimum rate.

{The second feature, depicted in~\FigRef{fig:proposedRelaxDeact}, is the ability to combine the relaxation of \ac{QoS} constraints concept and the user set reduction approach (cf. Section~\ref{Sec:RelatedWorks}) into a single QoS management mechanism.}
{In this example}, it is only possible to satisfy the minimum \ac{QoS} requirement of four out of the six users in the system.
The solution found by the proposed mechanism is a combination of the \ac{QoS} relaxation and user set reduction approaches, where the \ac{QoS} requirement of user-2 is relaxed while user-1 is not scheduled.
Such a result illustrates a situation where allocating some rate to user-1 would damage the \ac{QoS} of most remaining users, thus it is more efficient to not schedule user-1 and satisfy the \ac{QoS} demands of most remaining users.
Existing user set reduction-based solutions would deactivate two users (possibly user-1 and user-2) to meet the \ac{QoS} demands of the four remaining users, while existing mechanisms based on the relaxation of \ac{QoS} requirements would need to dramatically reduce the \ac{QoS} requirements of all users.
Therefore, the rate allocation in~\FigRef{fig:proposedRelaxDeact} illustrates the capabilities of the proposed solution of managing \ac{QoS} in an effective way to minimize the \ac{QoS} degradation suffered by the users.

\section{Algorithmic and Signaling Aspects {for 5G and Beyond Deployments}}

This section discusses how the proposed decentralized solution can be realized in practical 
{5G and beyond systems that operate as} \ac{MIMO}~\ac{IBC} systems.
Specifically, 
we discuss the tasks that {must} be executed by transmitters and receivers so that a decentralized \ac{QoS} control is achieved.
Moreover, we discuss the signaling aspects involved in a practical deployment of the proposed solution,
which enables our solution to use only local \ac{CSI} and run in a completely decentralized fashion.

\begin{figure}[!b]
	\centering
	\subfloat[Contemporary MIMO IBC scenario.]{\includegraphics[trim={2cm 1.5cm 1.9cm 1.3cm},clip,scale=0.6]{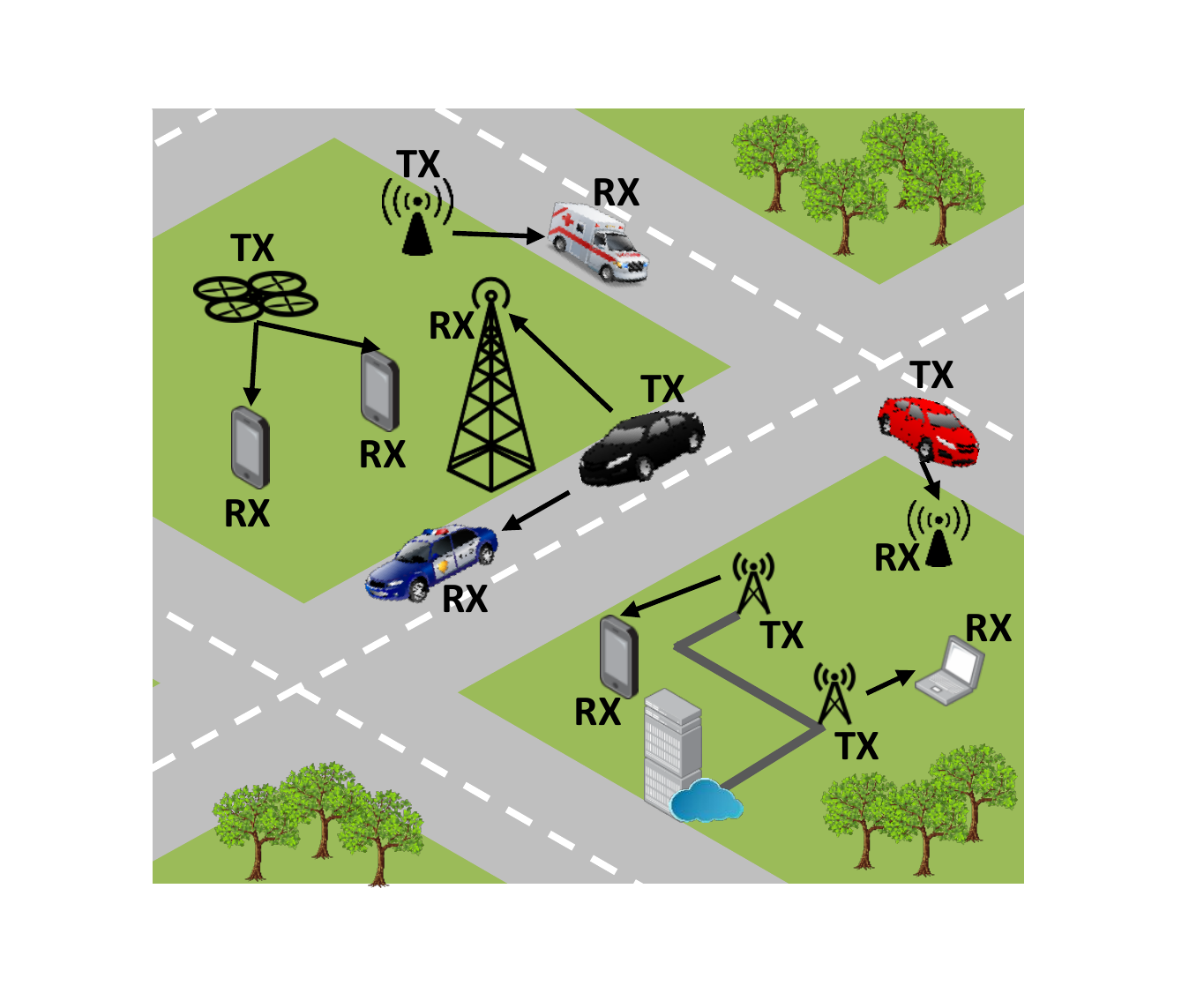}%
		\label{fig:Scenario}}
	
	\subfloat[Procedure executed by each TX.]{\includegraphics[scale=0.45]{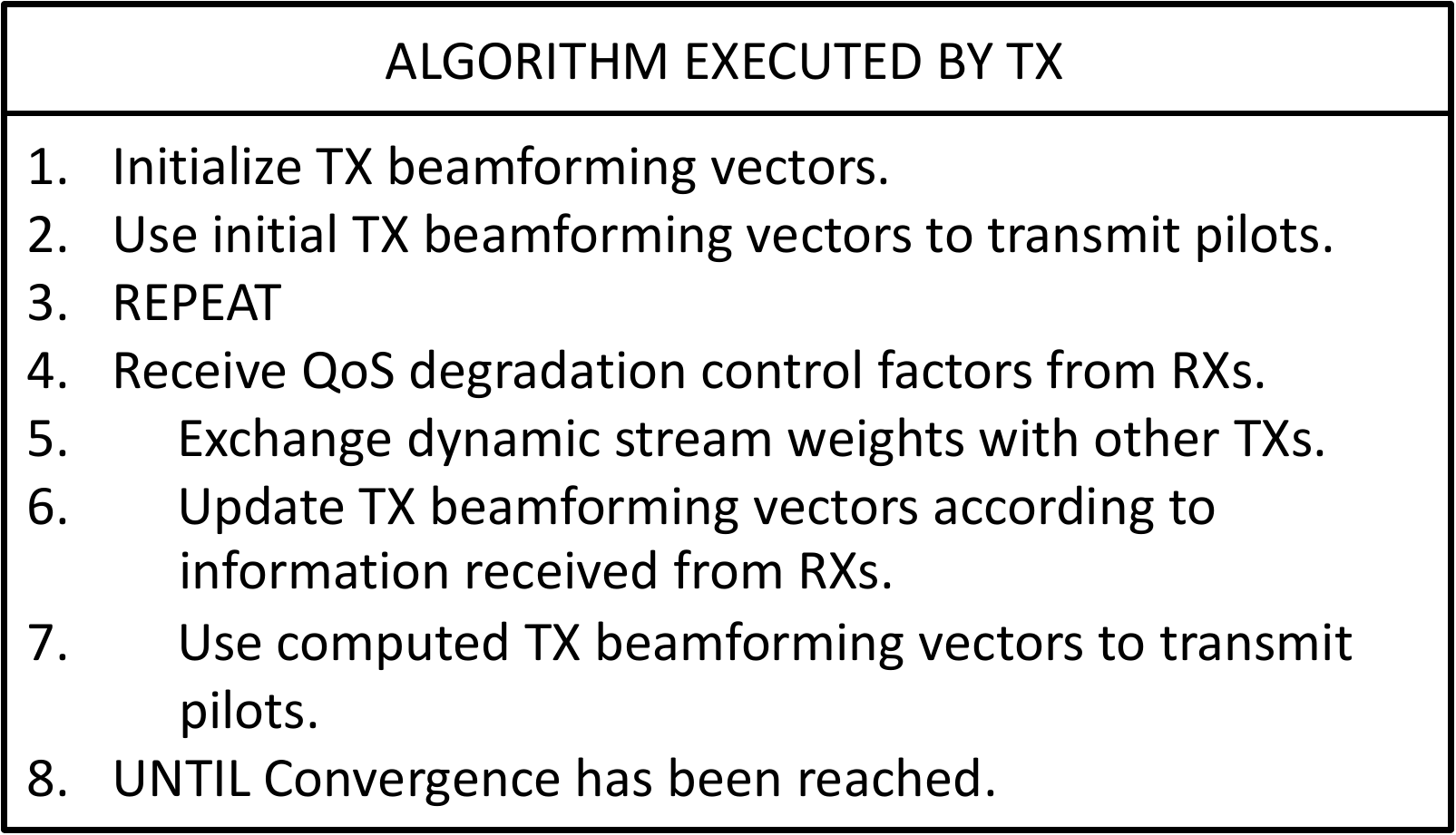}%
		\label{fig:TXAlg}}
	
	\subfloat[Procedure executed by each RX.]{\includegraphics[scale=0.45]{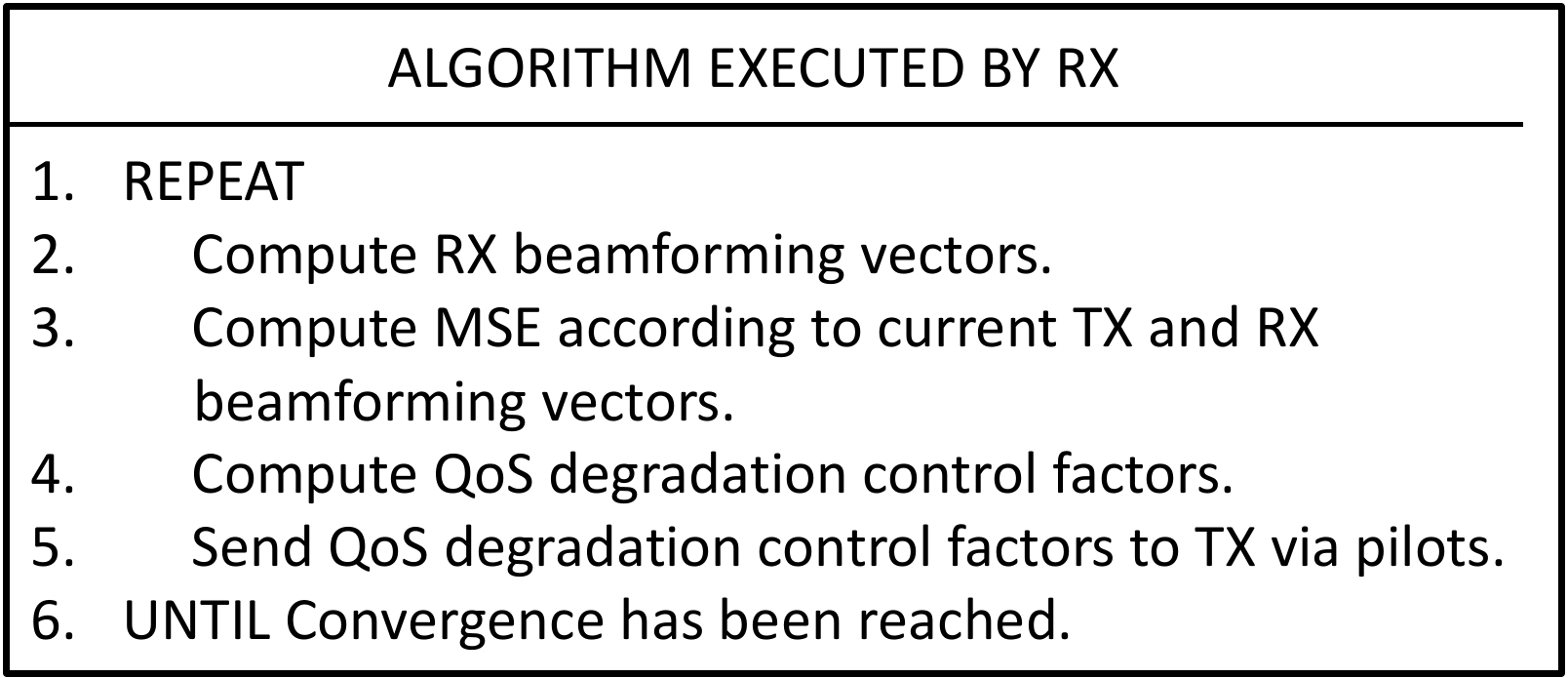}%
		\label{fig:RXAlg}}	
	
	\subfloat[Supporting frame structure.]{\includegraphics[scale=0.42]{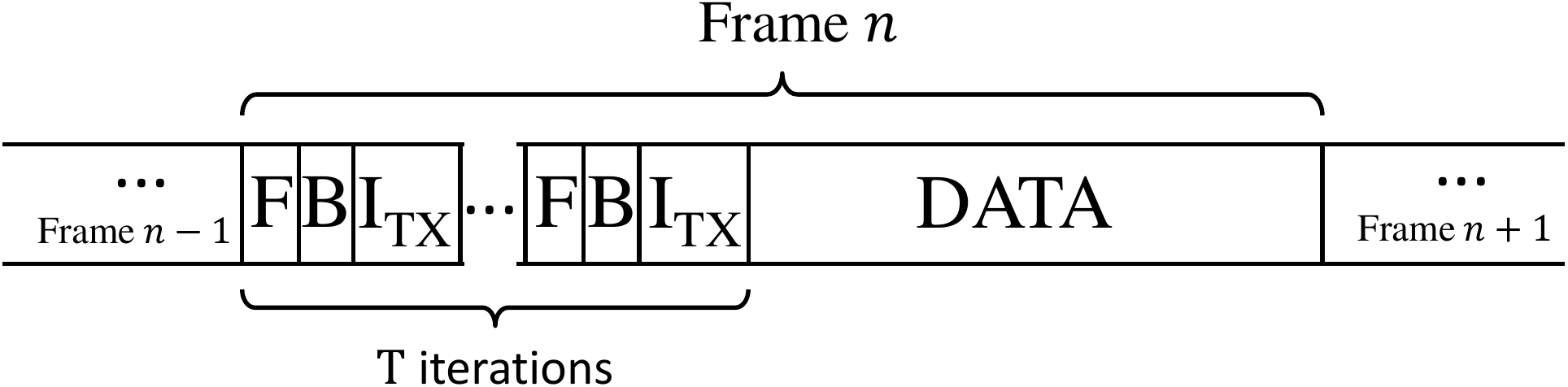}%
		\label{fig:FrameStrcuture}}	
	
	\caption{Practical aspects of proposed decentralized QoS degradation control for MIMO IBC systems.}
	\label{fig:PracticalAspects}
\end{figure}

\FigRef{fig:Scenario} illustrates a contemporary 
{5G scenario.} 
Such a scenario {is suitably modelled as a \ac{MIMO}~\ac{IBC} since it incorporates}
a variety of multi-antenna transmitters (TXs) and receivers (RXs).
The arrows indicate {the RX to which} each TX intends to send some data, where each TX can communicate with one or multiple RXs.
Considering that the TX-RX links {impose} some \ac{QoS} requirement, \FigRef{fig:Scenario}
shows a typical interference-limited \ac{MIMO} \ac{IBC} scenario where a decentralized \ac{QoS} degradation control is required.
In this {general setting},
the proposed solution can be deployed by means of each TX and RX executing
the iterative algorithm presented in \FigRef{fig:TXAlg} and \FigRef{fig:RXAlg}, respectively.
Those algorithms can be realized in practice using the frame structure depicted in~\FigRef{fig:FrameStrcuture},
which is based on~\cite{Tolli2019}, and extended to support the proposed solution.
{It is worth highlighting that the proposed frame structure is compatible with the \ac{3GPP} \ac{NR} specification, as detailed in~\cite{Tolli2019}.}

The proposed decentralized scheme is initiated by each TX computing
the respective initial TX beamforming vectors (step~1 in~\FigRef{fig:TXAlg})
using the maximum ratio transmission approach, for example.
{Next}, each TX transmits precoded pilots using the initial values of TX beamforming vectors (step~2 in~\FigRef{fig:TXAlg}).
These precoded pilots are used by each RX to estimate their local \ac{CSI} and \ac{MSE}
as well as to compute the respective RX beamforming vectors, which are steps 2 and 3 in~\FigRef{fig:RXAlg}.
The next steps performed by each RX are the computation of the \ac{QoS} degradation control factors and
the transmission of those values via precoded pilots to the respective TXs, which encompasses steps 4 and 5 in~\FigRef{fig:RXAlg}.
Then, each TX receives the information transmitted by its intended RXs in step~4 of~\FigRef{fig:TXAlg}.
After that, each TX exchanges dynamic stream weights with its neighboring TXs in step~4 of~\FigRef{fig:TXAlg},
which are used to control how much rate should be allocated to each RX
in order to meet or minimally degrade the \ac{QoS} demands.
Finally, each TX updates its respective TX beamforming vectors and
use them to transmit precoded pilots in steps 6 and 7 in~\FigRef{fig:TXAlg}, respectively.
All the described steps comprise one iteration of the proposed iterative solution.

The proposed frame structure is divided into two main parts:
one used for the transceiver beamforming vectors configuration,
while the second comprises the actual data transmission.
In the beamforming configuration phase,
the over-the-air signaling is split into two parts: the letter \textbf{F}
denotes the forward pilot transmission, which occurs from the TXs to the RXs
and corresponds to steps 2 and 7 in~\FigRef{fig:TXAlg}, while the letter \textbf{B}
denotes the backward pilot transmission, which happens
from each RX to its serving TX and corresponds to step 5 in~\FigRef{fig:RXAlg}.
Finally, $\textbf{I}_\textbf{TX}$ denotes the signaling that occurs between TXs,
which is used to share the dynamic stream weights in step 5 in~\FigRef{fig:TXAlg}.
When available, a backhaul link can be used in the $\textbf{I}_\textbf{TX}$,
otherwise over-the-air signaling should be employed.

{The computational complexity of algorithms derived from problem~\eqref{PROB:MAX_RATE_SLACK} when using a \ac{SCA} approach to solve the problem of penalized sum-rate maximization with minimum rate constraints is in the order of $\mathcal{\uppercase{O}}(U^2N_TN_R^2 + U^2N_T^2N_R + U^2N_T^3 + UN_R^3)$. Such a computational complexity comes from the computation of the transmit beamforming vector, the \ac{MMSE} receivers and \ac{MSE} values. For different objective functions and types of \ac{QoS} requirement, the computational complexity of the derived algorithms need to be further analyzed, which is out of the scope of this work.}

\section{Performance Evaluation}

This section analyses the performance of the proposed solution considering the \ac{MIMO} \ac{IBC} scenarios shown in~\FigRef{fig:intro:scenarios} and~\FigRef{fig:Scenario}.
These scenarios are more challenging than existing cellular \ac{MIMO} \ac{IBC} systems due to the random positioning of both transmitters and receivers, which creates complicated interference patterns.

The simulated scenarios consider that 10 transmitters and the respective 2 intended receivers of each transmitter are randomly dropped within a 400~m radius region.
Transmitters and receivers are equipped with 8 and 4 antennas each, respectively, which represents a possible contemporary \ac{MIMO}~\ac{IBC} scenario where the devices {do not employ} massive \ac{MIMO}{, such as the scenarios in~\cite{3gpp.38.885} and~\cite{Petrov2019}}.
The channel matrices are computed using the \ac{WINNER} II B1 channel model.
The per-transmitter power budget is 35~dBm.
{The results are averaged over 400 Monte-Carlo simulations.}
The objective of the proposed solution is {maximizing} the total system rate subject to per-link minimum rate constraints.
For performance comparison, we use the \ac{WMMSE} algorithm~\cite{Shi2011} and the algorithm in~\cite{Kaleva2016}, which are also decentralized algorithms and have similar computational complexity {as} the proposed solution.

\begin{figure}[!b]
	\centering	
	\subfloat[Average values of QoS degradation.]{\includegraphics[width=0.9\columnwidth]{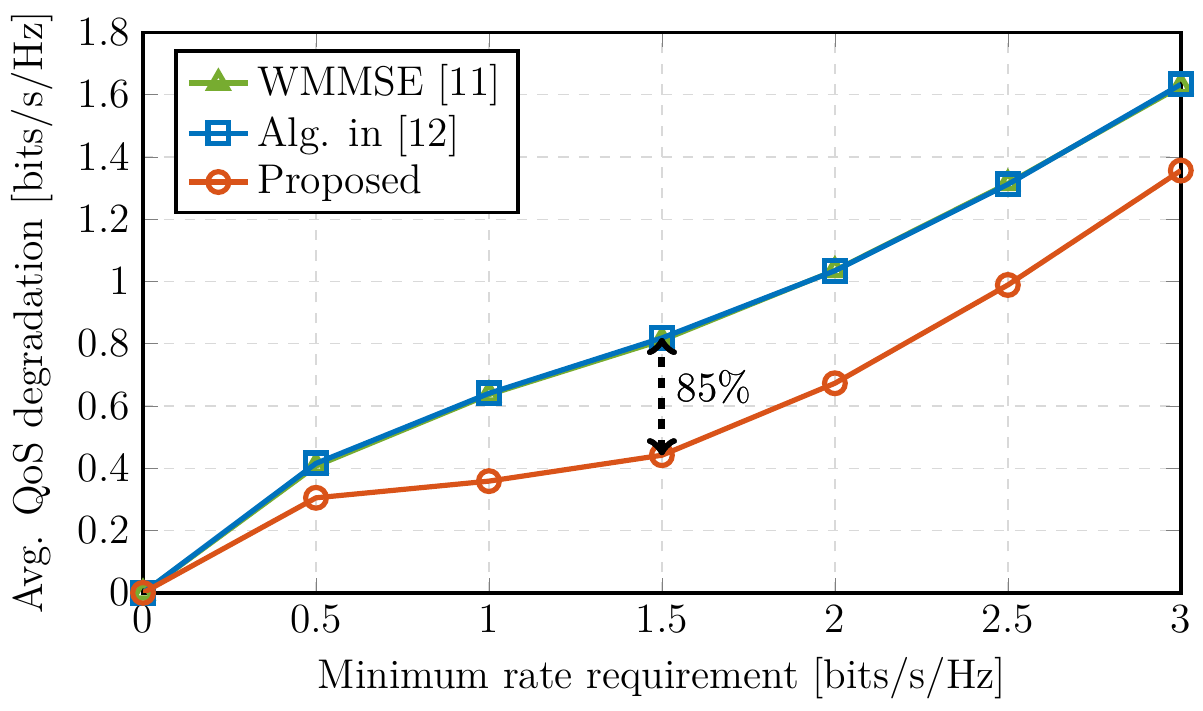}%
		\label{fig:avgslack}}

	\subfloat[CDF of QoS degradation values.]{\includegraphics[width=0.9\columnwidth]{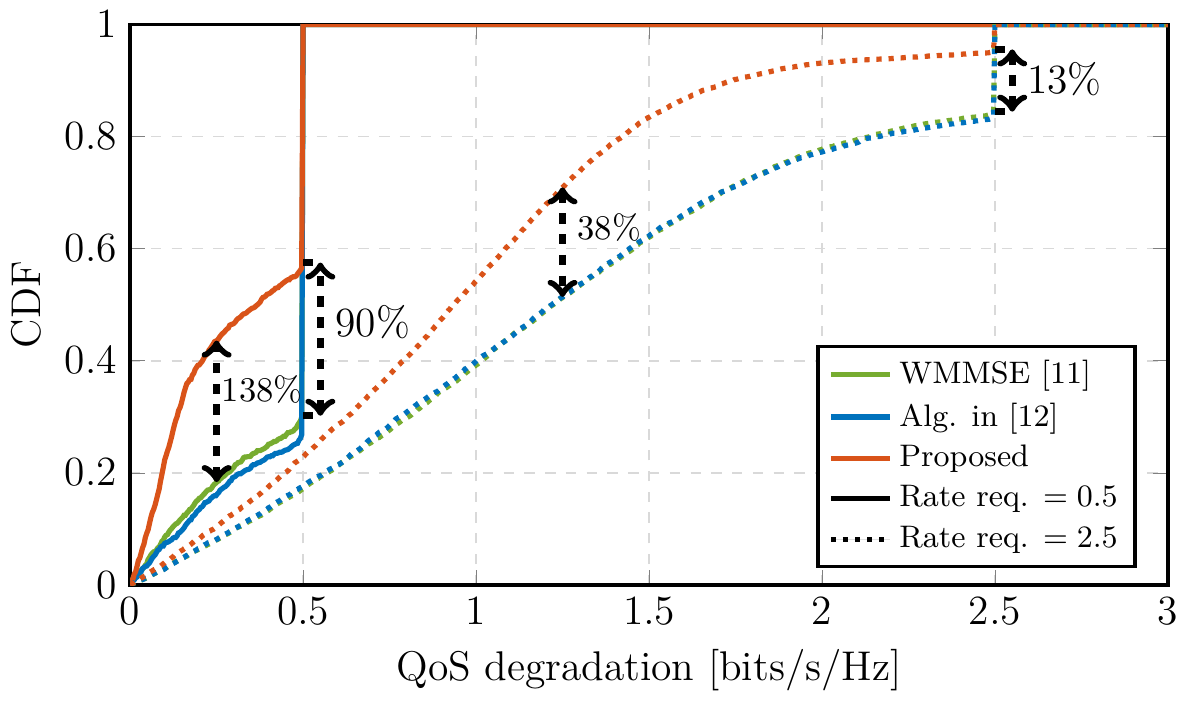}%
		\label{fig:cdfOfSlacks}}	
	\caption{Analysis of \ac{QoS} degradation values in contemporary \ac{MIMO} \ac{IBC} scenarios, showing that the proposed joint transceiver design and link scheduling
		for \ac{QoS} management effectively provides an enhanced \ac{QoS} control.}
	\label{fig:peformanceEval}
\end{figure}

\FigRef{fig:avgslack} shows the average \ac{QoS} degradation when increasing the minimum rate requirement, where we consider only the links that {do not meet the original rate requirements}.
The \ac{QoS} degradation is the difference between the rate achieved by each link and its minimum rate requirement, thus it is given in the unit of rate.
For the rate-unconstrained scenario, i.e., for the rate requirement of $0$ bits/s/Hz, all algorithms present the same performance.
This occurs because, in this case, there is no need for any link scheduling, so the proposed algorithm and the algorithm in~\cite{Kaleva2016}, which consider rate requirements in their formulations, reduce to the rate-unconstrained case dealt by the \ac{WMMSE} algorithm.
When the rate requirement increases, the proposed solution achieves gains up to 85\% over both comparison algorithms, while presenting a very small performance loss up to 5\% in terms of total system rate.
On the other hand, the algorithm in~\cite{Kaleva2016} and the \ac{WMMSE} algorithm present very similar performances.
This shows that, in terms of average \ac{QoS} degradation, algorithms that consider minimum \ac{QoS} requirements but do not take into account any link scheduling mechanism ({e.g.}, the solution in~\cite{Kaleva2016}) perform as {poorly} as {algorithms} that do not handle \ac{QoS} demands ({e.g.}, the \ac{WMMSE} algorithm) in \ac{MIMO}~\ac{IBC} scenarios.
This shows the importance of a complete solution that jointly handles {transceiver design, link scheduling and \ac{QoS} management}, such as the proposed solution.

\FigRef{fig:cdfOfSlacks} shows the \ac{CDF} of the \ac{QoS} degradation values for the links that did not meet the respective rate requirements.
This is shown for two values of minimum rate requirement, namely, $0.5$ and $2.5$ bits/s/Hz.
{Notice that} in terms of user satisfaction, the algorithm in~\cite{Kaleva2016} performs similarly to the \ac{WMMSE} algorithm, which is designed for rate-unconstrained scenarios and, as such, does not deal with QoS demands.
For the rate requirement of $0.5$~bits/s/Hz, the proposed solution achieves higher user satisfaction in two ways: (a) unsatisfied links are consistently scheduled with rates closer to their minimum rate requirements, where the proposed solution achieves a gain of 138\% considering unsatisfied links that get at least half of their rate requirement; and (b) the proposed solution provides a 90\% gain in terms of deactivated links, i.e., links that are allocated with zero rate.
For a higher value of minimum rate requirement, those gains are as high as 38\% and 13\%, respectively.
This gain reduction occurs because in highly demanding scenarios, i.e., when the minimum rate requirement is further increased, the proposed solution has less freedom due to the reduced set of transceiver beamforming vectors that allow for a smaller \ac{QoS} degradation with respect to the comparison algorithms.
Nevertheless, the presented results show that designing a complete solution that jointly handles the transceiver design and link scheduling
for \ac{QoS} management provides significant gains in contemporary \ac{MIMO} \ac{IBC} scenarios.

\section{Conclusions {and Outlook}}
Many current and emerging multi-antenna scenarios {currently studied by the 3GPP community},
{including massive and broadband Internet of Things applications},
can be characterized as \ac{MIMO}~\ac{IBC} systems,
which require decentralized mechanisms {for transceiver design, link scheduling, and \ac{QoS} management}.
Starting from a survey of the major works in this research field,
this work  proposed a {solution concept for joint and distributed}
transceiver design and link scheduling {with} \ac{QoS} management to fully harvest the potential gains in such systems.
A {key} feature of our solution is the introduction of a QoS management mechanism
that allows and optimizes, when necessary, a  degradation of \ac{QoS} per individual users
while minimizing the overall users' dissatisfaction in the system.  
Algorithmic and signaling aspects were also presented for practical implementations of the proposed solution.
Numerical results demonstrated that our solution provides
an enhanced \ac{QoS} degradation control compared to existing solutions,
proving that handling the user scheduling and transceiver design
in a decentralized fashion in {5G and beyond wireless communications systems operating in the} \ac{MIMO} \ac{IBC}
is feasible and beneficial for managing \ac{QoS} in a resource-efficient manner.

{This work also provided a high-level description of the first steps toward a real implementation. 
Due to recent developments in \ac{3GPP} related to advanced \ac{QoS} management,
\ac{HARQ} mechanisms, channel state information acquisition, and resource allocation for both the cellular (Uu) 
and the sidelink (PC5) interfaces, including multihop sidelink relaying, and integrated access backhaul (IAB) signaling,
the standardization community has already made the first steps towards decentralized \ac{MIMO} \ac{IBC} implementations supporting
flexible topologies~\cite{Ashraf2020}. Therefore, this work connects the near-optimal theoretical solution to algorithmic and high-level implementation aspects.}

\bibliographystyle{IEEEtran}
\bibliography{IEEEabrv,biblio}

\section*{Biographies}
{\footnotesize
	\begin{itemize}
		\item \textbf{Roberto Pinto Antonioli} received his B.Sc. degree in teleinformatics engineering (magna cum laude) from the Federal University of Cear\'a (UFC), Fortaleza, Brazil, in 2016. He received his M.Sc. and PhD degrees in teleinformatics engineering also from UFC in 2017 and 2020, respectively. He currently holds a post-doc position at the Wireless Telecom Research Group (GTEL), UFC, where he works on projects in technical and scientific cooperation with Ericsson Research. In 2018/2019, he was a visiting researcher at Ericsson Research in Sweden. His research interests include 5G wireless communication networks with multiple radio access technologies and multi-connectivity, as well as scheduling algorithms for QoS provision.
	\end{itemize}
	\begin{itemize}
		\item \textbf{G\'abor Fodor} [SM'08] received the Ph.D. degree in electrical engineering from the Budapest University of Technology and Economics in 1998, and the D.Sc. degree of the Hungarian Academy of Sciences (doctor of MTA) in 2019. He is currently a master researcher at Ericsson Research and a docent and adjunct professor at KTH Royal Institute of Technology, Stockholm, Sweden. He has authored or co-authored more than 100 refereed journal and conference papers, seven book chapters and more than 100 European and US granted patents. He was a co-recipient of the IEEE Communications Society Stephen O. Rice prize in 2018 and the Best Student Conference Paper award by the IEEE Sweden VT/COM/IT Chapter in 2018. Dr. Fodor is currently the chair of the IEEE Communications Society Emerging Technology Initiative on Full Duplex Communications. Between 2017-2020 was also a member of the board of the IEEE Sweden joint Communications, Information Theory and Vehicle Technology chapter. He is currently serving as an Editor of the IEEE Transactions on Wireless Communications, a Guest Editor for the IEEE Communications Magazine (Special Issue on Terahertz communications) and a Guest Editor for the IEEE Wireless Communications (Special Issue on full-duplex communications).
	\end{itemize}
	\begin{itemize}
		\item \textbf{Pablo Soldati} received his M.Sc. and Ph.D. degrees in wireless communications from the University of Siena, Italy, and Royal Institute of Technology KTH in 2004 and 2010, respectively. He then was a postdoctoral scholar at KTH and visiting postdoctoral researcher at Stanford University. In June 2011 he joined Huawei Technologies Sweden AB as Senior Researcher for advanced radio transmission technology and standards, and regularly participated as Huawei's delegate to 3GPP RAN1 meetings for LTE-A standardization. From 2014 he held a Principal Researcher position with the System Algorithms group at Huawei Technologies Sweden AB working on advanced radio resource management algorithms for 5G mobile broadband systems based on machine learning and artificial intelligence. In 2018 he joined Ericsson AB, where he currently serves as a standardization and concepts researcher, contributing to the 3GPP RAN3 5G standardization and to development of Ericsson RAN solutions based on artificial intelligence. His research interests include non-convex and distributed optimization, radio access networks, machine learning and artificial intelligence. More details are available at http://www.pablosoldati.com/
	\end{itemize}
	\begin{itemize}
		\item \textbf{Tarcisio Ferreira Maciel} received his B.Sc. and M.Sc. degrees in Electrical Engineering from the Federal University of Cear\'a (UFC) in 2002 and 2004, respectively, and his Dr.-Ing. degree from the Technische Universitat Darmstadt (TUD), Germany, in 2008, also in Electrical Engineering. Since 2001, he has actively participated in several projects in a technical and scientific cooperation between the Wireless Telecommunications Research Group (GTEL), UFC, and Ericsson Research. From 2005 to 2008, he was a research assistant with the Communications Engineering Laboratory, TUD. Since 2008, he has been a member of the Post-Graduation Program in Teleinformatics Enginnering, UFC. In 2009, he was a professor of computer engineering with UFC-Sobral. From 2010 to 2015, he was a professor with the Center of Technology, UFC. Since 2015, he is an associate professor with the Teleinformatics Engineering Department (DETI), UFC. His research interests include radio resource allocation, optimization, and multiuser/multiantenna communications.
\end{itemize}}

\end{document}